\documentclass[submission,copyright,creativecommons]{eptcs}
\providecommand{\event}{GandALF 2018} % Name of the event you are submitting to
\usepackage{breakurl}             % Not needed if you use pdflatex only.
\usepackage{underscore}           % Only needed if you use pdflatex.

\usepackage{xspace}
\usepackage[utf8]{inputenc}
\usepackage{enumitem}

\usepackage{amsmath}
\usepackage{amsfonts}
\usepackage{amssymb}
\usepackage{amsthm}
\usepackage{xstring}
\usepackage{url}
\usepackage{tabularx}
\usepackage{xspace}
\usepackage{hyperref}

\usepackage{tikz}
\usetikzlibrary{decorations.pathreplacing}
\usetikzlibrary{decorations.pathmorphing}
\usetikzlibrary{shapes,shapes.symbols,automata,arrows}
\usetikzlibrary{calc}
\usetikzlibrary{patterns}

\tikzset{p0/.style = {shape = circle, draw, thick, minimum size = 0.7cm}}
\tikzset{p1/.style = {rectangle, minimum size=.7cm, draw, thick}}
\tikzset{>=stealth, shorten >=1pt}
\tikzset{every edge/.style = {thick, ->, draw}}
\tikzset{every loop/.style = {thick, ->, draw}}

\makeatletter
\tikzset{circle split part fill/.style  args={#1,#2}{%
 alias=tmp@name, % Jake's idea !!
  postaction={%
    insert path={
     \pgfextra{% 
     \pgfpointdiff{\pgfpointanchor{\pgf@node@name}{center}}%
                  {\pgfpointanchor{\pgf@node@name}{east}}%            
     \pgfmathsetmacro\insiderad{\pgf@x}
      \fill[#1] (\pgf@node@name.base) ([xshift=-\pgflinewidth]\pgf@node@name.east) arc
                          (0:180:\insiderad-\pgflinewidth)--cycle;
      \fill[#2] (\pgf@node@name.base) ([xshift=\pgflinewidth]\pgf@node@name.west)  arc
                           (180:360:\insiderad-\pgflinewidth)--cycle; 
         }}}}}  
 \makeatother

\usepackage{booktabs}

\usepackage{changebar}

\theoremstyle{plain}
\newtheorem{thm}{Theorem}
\newtheorem{lem}[thm]{Lemma}
\newtheorem{rem}[thm]{Remark}
\newtheorem{cor}[thm]{Corollary}

\newtheorem{prop}[thm]{Proposition}

\usepackage{pifont}

\definecolor{myred}{rgb}{1,0.604,0.604}
\definecolor{mydarkred}{rgb}{1,0.345,0.345}
\definecolor{myblue}{rgb}{0.635,0.675,0.966}
\definecolor{mydarkblue}{rgb}{0.412,0.475,0.957}
\definecolor{myyellow}{rgb}{1,0.976,0.604}
\definecolor{mydarkyellow}{rgb}{1,0.961,0.345}

\tikzset{
	assign/.style = { fill=myblue },
	choice/.style = { fill=myred },
	check/.style  = { fill=myyellow }
}

\makeatletter
\DeclareRobustCommand{\rvdots}{%
  \vbox{
    \baselineskip4\p@\lineskiplimit\z@
    \kern-\p@
    \hbox{.}\hbox{.}\hbox{.}
  }}
\makeatother

\newcommand\restr[2]{{% we make the whole thing an ordinary symbol
  \left.\kern-\nulldelimiterspace % automatically resize the bar with \right
  #1 % the function
  \vphantom{\big|} % pretend it's a little taller at normal size
  \right|_{#2} % this is the delimiter
  }}

\newcommand{\ext}{\mathrm{ext}}

\newcommand{\rank}{\mathrm{rk}}

\newcommand{\subarenaof}{\sqsubseteq}

\newcommand{\initmark}{I}

\newcommand{\vinit}{v_{\initmark}}

\newcommand{\myquot}[1]{``#1''}

\newcommand{\bigo}[0]{\mathcal{O}}

\newcommand{\size}[1]{|#1|}
\newcommand{\card}[1]{\size{#1}}
\newcommand{\set}[1]{\{ #1 \}}
\newcommand{\nats}{\mathbb{N}}

\renewcommand{\epsilon}{\varepsilon}

\newcommand{\arena}{\mathcal{A}}
\newcommand{\game}{\mathcal{G}}
\newcommand{\gamelim}{\mathcal{G}_{\lim}}
\newcommand{\gamesup}{\mathcal{G}_{\sup}}
\newcommand{\cost}{\mathrm{Cost}}

\newcommand{\wincond}{\mathrm{Win}}

\newcommand{\val}{\mathrm{val}}

\newcommand{\reducesto}{\leq}

\newcommand{\safety}{\mathrm{Safety}}

\newcommand{\mem}{\mathcal{M}}
\newcommand{\init}{m_\initmark}
\newcommand{\update}{\mathrm{Upd}}
\newcommand{\nxt}{\mathrm{Nxt}}

\newcommand{\reqres}{\textsc{ReqRes}}
\newcommand{\costreqres}{\textsc{CostReqRes}}
\newcommand{\reqresdist}{\textsc{ReqResCor}}

\newcommand{\att}[3]{\mathrm{Attr}_{#1}^{#2}(#3)}

\newcommand{\exptime}{\textsc{ExpTime}}

\newcommand{\pspace}{\textsc{PSpace}\xspace}

\title{Quantitative Reductions and Vertex-Ranked Infinite Games\thanks{Supported by by the project ``TriCS'' (ZI 1516/1-1) of the German Research Foundation (DFG) and the Saarbrücken Graduate School of Computer Science}}
\author{Alexander Weinert
\institute{Reactive Systems Group, Saarland University, 66123 Saarbr{\"u}cken, Germany}
\email{weinert@react.uni-saarland.de}
}
\def\titlerunning{Quantitative Reductions and Vertex-Ranked Infinite Games}
\def\authorrunning{A. Weinert}
\begin{document}
\maketitle

\begin{abstract}
	We introduce quantitative reductions, a novel technique for structuring the space of quantitative games and solving them that does not rely on a reduction to qualitative games.
	We show that such reductions exhibit the same desirable properties as their qualitative counterparts and additionally retain the optimality of solutions.
	Moreover, we introduce vertex-ranked games as a general-purpose target for quantitative reductions and show how to solve them.
	In such games, the value of a play is determined only by a qualitative winning condition and a ranking of the vertices.
	
	We provide quantitative reductions of quantitative request-response games to vertex-ranked games, thus showing \exptime-completeness of solving the former games.
	Furthermore, we exhibit the usefulness and flexibility of vertex-ranked games by showing how to use such games to compute fault-resilient strategies for safety specifications.
	This work lays the foundation for a general study of fault-resilient strategies for more complex winning conditions.
\end{abstract}

\section{Introduction}
\label{sec:introduction}

The study of quantitative infinite games has garnered great interest lately, as they allow for a much more fine-grained analysis and specification of reactive systems than classical qualitative games~\cite{BrimCDGR11,BruyereFiliotRandourRaskin17,BruyereHautemRandour16,EhrenfeuchtMycielski79,FZ14,WeinertZimmermann16,ZwickPaterson95}.
While there exists previous work investigating quantitative games, the approaches to solving them usually rely on ad-hoc solutions that are tailor-made to the considered winning condition.
Moreover, quantitative games are usually solved by reducing them to a qualitative game in a first step, hardcoding a certain value of interest during the reduction.
In particular, to the best of our knowledge, no general framework for the analysis of such games that is analogous to the existing one for qualitative games has been developed.
In this work, we introduce such a framework that disentangles the study of quantitative games from that of qualitative ones.

Qualitative infinite games have been applied successfully in the verification and synthesis of reactive systems~\cite{BloemChatterjeeEtAl14,BloemChatterjeeHenzingerJobstmann09,BouyerMarkeyOlschewskiUmmels11,CernyChatterjeeHenzingerRadhakrishnaSingh11}.
They have given rise to a multitude of algorithms that ascertain system correctness and that synthesize correct-by-construction systems.
In such a game, two players, called Player~$0$ and Player~$1$, move a token in a directed graph.
After infinitely many moves, the resulting sequence of vertices is evaluated and one player is declared the winner of the play.
For example, in a qualitative request-response game~\cite{WallmeierHuettenThomas03}, the goal for Player~$0$ is to ensure that every visit to a vertex denoting some request is eventually followed by a visit to a vertex denoting an answer to that request.
In order to solve qualitative games, i.e., to determine a winning strategy for one player, one often reduces a complex game to a potentially larger, but conceptually simpler one.
For example, in a multi-dimensional request-response game, i.e., in a request-response game in which there exist multiple conditions that can be requested and answered, one stores the set of open requests and demands that every request is closed at infinitely many positions.
As this is a Büchi condition, which is much simpler than the request-response condition, one is able to reduce request-response games to Büchi games.

In recent years the focus of research has shifted from the study of qualitative games, in which one player is declared the winner of a given play, to that of quantitative games, in which the resulting play is assigned some value or cost.
Such games allow modeling systems in which, for example, requests have to be answered within a certain number of steps~\cite{BruyereHautemRandour16,ChatterjeeHenzingerHorn09,FaymonvilleZimmermann14,FZ14,KupfermanPitermanVardi09}, systems with one or more finite resources which may be drained and charged~\cite{BouyerMarkeyRandourLarsenLaursen16,ChatterjeeDoyen12,ChatterjeeDoyenHenzingerRaskin10,DBLP:journals/corr/abs-1209-3234}, or scenarios in which each move incurs a certain cost for either player~\cite{EhrenfeuchtMycielski79,ZwickPaterson95}.

In general, Player~$0$ aims to minimize the cost of the resulting play, i.e., to maximize its value, while Player~$1$ seeks to maximize the cost, thus minimizing the value.
In a quantitative request-response game, for example, it is the goal of Player~$0$ to minimize the number of steps between requests and their corresponding answers.
The typical questions asked in the context of such games are
\myquot{Does there exist an upper bound on the time between requests and responses that Player~$0$ can ensure?}~\cite{ChatterjeeHenzingerHorn09,FZ14,KupfermanPitermanVardi09,ScheweWeinertZimmermann18},
\myquot{Can Player~$0$ ensure an average cost per step greater than zero?}~\cite{ZwickPaterson95},
\myquot{What is the minimal time between requests and responses that Player~$0$ can ensure?}~\cite{WeinertZimmermann16}, or
\myquot{What is the minimal average level of the resource that Player~$0$ can ensure without it ever running out?}~\cite{BouyerMarkeyRandourLarsenLaursen16}.
The former two questions can be seen as boundedness questions, while the latter two are asking for optimal solutions.

Such decision problems are usually solved by fixing some bound~$b$ on the cost of the resulting plays and subsequently reducing the quantitative game to a qualitative one, hardcoding the fixed~$b$ in the process.
If the cost of the resulting play is below~$b$, then Player~$0$ is declared the winner.
For example, in order to determine the winner in a quantitative request-response game as described above for some bound~$b$, we construct a Büchi game in which every time a request is opened, a counter for that request is started which counts up to the bound~$b$ and is reset if the request is answered.
Once any counter exceeds the value~$b$, we move to a terminal position indicating that Player~$0$ has lost.
We then require that every counter is inactive infinitely often, which is again a Büchi condition and thus much simpler than the original quantitative request-response condition.
Player~$0$ wins the resulting game if and only if she can ensure that every request is answered within at most~$b$ steps.

Such reductions are usually specific to the problem being addressed.
Furthermore, they immediately abandon the quantitative aspect of the game under consideration, as the bound is hardcoded during the first step of the analysis.
Thus, even when only changing the bound one is interested in, the reduction has to be recomputed and the resulting qualitative game has to be solved from scratch.
In our request-response example, if one is interested in checking whether Player~$0$ can ensure every request to be answered within at most~$b' \neq b$ steps, one constructs a new Büchi game for the bound~$b'$.
This game is then solved independently of the one previously computed for the bound~$b$.

In this work, we lift the concept of reductions for qualitative games to quantitative games.
Such quantitative reductions enable the study of a multitude of optimization problems for quantitative games in a way similar to decision problems for qualitative games.
When investigating quantitative request-response games using quantitative reductions, for example, we only compute a single, simpler quantitative game and subsequently check this game for a winning strategy for Player~$0$ for any bound~$b$.
If she has such a strategy in the latter game, the quantitative reduction yields a strategy for her satisfying the same bound in the former one.

In general, we retain the intuitive property of reductions for qualitative games:
The properties of a complex quantitative game can be studied by investigating a potentially larger, but conceptually simpler quantitative game.

\subparagraph*{Contributions}
\label{sec:introduction:contributions}
We present the first framework for reductions between quantitative games and we provide vertex-ranked games as general-purpose targets for such reductions.
Moreover, we show tight bounds on the complexity of solving vertex-ranked games with respect to a given bound.

Finally, we provide two examples illustrating the use of the concepts introduced in this work:
First, we define quantitative request-response games and solve them using quantitative reductions to vertex-ranked games.
Second, we illustrate the versatility of vertex-ranked games by using them to compute fault-resilient strategies for safety games with faults.

After introducing qualitative and quantitative games formally in Section~\ref{sec:preliminaries}, we define quantitative reductions in Section~\ref{sec:quantitative-reductions} and show that they provide a mechanism to solve a quantitative game with respect to a given bound:
If a game~$\game$ can be reduced to a game~$\game'$, then we can use a strategy for Player~$0$ that minimizes the cost of plays in~$\game'$ to construct a strategy for her which minimizes the cost of plays in~$\game$.

In Section~\ref{sec:vertex-ranked-games}, we define vertex-ranked games, very general classes of quantitative games that can be used as targets for quantitative reductions.
The quantitative condition of such games is quite simple, as the cost of a play is determined only by a qualitative winning condition and a ranking of the vertices of the game.
If the resulting play is winning according to the qualitative condition, then its cost is given by the highest rank visited at all or visited infinitely often, depending on the particular variant of vertex-ranked games.
Otherwise, the value of the play is infinite.
We show that solving such vertex-ranked games is as hard as solving the underlying qualitative winning condition.

Finally, in Section~\ref{sec:application} we provide two examples of the versatility of vertex-ranked games:
First, we define and solve request-response games with costs via quantitative reductions.
Second, we discuss how to use vertex-ranked games to compute fault-resilient strategies in safety games with faults.
In such games, after Player~$0$ has picked a move, say to vertex~$v$, a fault may occur, which overrides the choice of Player~$0$ and the game continues in vertex~$v' \neq v$ instead.
By using vertex-ranked games, we are able to compute strategies that are resilient against as many faults as possible.

Proofs omitted due to space restrictions as well as an additional example can be found in the full version~\cite{Weinert17}.

\section{Preliminaries}
\label{sec:preliminaries}

We first define notions that are common to both qualitative and quantitative games.
Afterwards, we recapitulate the standard notions for qualitative games before defining quantitative games and lifting the notions for qualitative games to the quantitative case.

We denote the non-negative integers by $\nats$ and define $[n] = \set{0, 1, \ldots, n-1}$ for every $n \ge 1$.
Also, we define~$\infty > n$ for all~$n \in \nats$ and~$\nats_\infty = \nats \cup \set{\infty}$.
An \textit{arena}~$\arena=(V, V_0, V_1, E, \vinit)$ consists of a finite, directed graph~$(V, E)$, a partition~$(V_0, V_1)$ of $V$ into the positions of Player~$0$ and Player~$1$, and an initial vertex~$\vinit \in V$. The size of $\arena$, denoted by $\size{\arena}$, is defined as $\size{V}$.
A \textit{play} in $\arena$ is an infinite path~$\rho = v_0 v_1 v_2 \cdots$ through $(V, E)$ starting in $\vinit$.
To rule out finite plays, we require every vertex to be non-terminal.

A \textit{strategy} for Player~$i$ is a mapping $\sigma \colon V^*V_i \rightarrow V$ that satisfies $(v, \sigma(\pi v)) \in E$ for all $\pi \in V^*$, $v \in V_i$.
We say that $\sigma$ is \textit{positional} if $\sigma(\pi v) = \sigma(v)$ for every $\pi \in V^*$, $v \in V_i$.
A play $v_0 v_1 v_2 \cdots$ is \textit{consistent} with a strategy~$\sigma$ for Player~$i$, if $v_{j+1} = \sigma( v_0 \cdots v_j)$ for all~$j$ with $v_j \in V_i$.

A \textit{memory structure}~$\mem = (M, \init, \update)$ for an arena $(V, V_0, V_1, E, \vinit)$ consists of a finite set~$M$ of memory states, an initial memory state $\init \in M$, and an update function~$\update\colon M \times V \rightarrow M$.
We extend the update function to finite play prefixes in the usual way:
$\update^+(v_\initmark) = m_\initmark$ and $\update^+(\pi v) = \update(\update^+(\pi), v)$ for play prefixes $\pi \in V^+$ and $v \in V$.
A next-move function $\nxt \colon V_i \times M \rightarrow V$ for Player~$i$ has to satisfy $(v, \nxt(v, m)) \in E$ for all $v \in V_i$, $m \in M$.
It induces a strategy~$\sigma$ for Player~$i$ with memory~$\mem$ via $\sigma(v_0\cdots v_j) = \nxt(v_j, \update^+(v_0 \cdots v_j))$.
A strategy is called \textit{finite-state} if it can be implemented by a memory structure.
We define $\card{\mem} = \card{M}$.
In a slight abuse of notation, the size~$\card{\sigma}$ of a finite-state strategy is the size of a memory structure implementing it.

An arena $\arena = (V, V_0, V_1, E, \vinit)$ and a memory structure $\mem = (M, \init, \update)$ for $\arena$ induce the expanded arena $\arena\times\mem = (V \times M, V_0 \times M, V_1 \times M, E', (\vinit, \init))$, where~$E'$ is defined via $((v,m), (v',m')) \in E'$ if and only if $(v,v') \in E$ and $\update(m, v' ) = m'$.
Every play $\rho = v_0 v_1 v_2\cdots$ in $\arena$ has a unique extended play $\ext_\mem(\rho) = (v_0, m_0) (v_1, m_1)
(v_2, m_2) \cdots$ in $\arena \times \mem$ defined by $m_0 = \init$ and $m_{j+1} = \update(m_j, v_{j+1})$, i.e., $m_j = \update^+(v_0 \cdots v_j)$.
We omit the index~$\mem$ if it is clear from the context.
The extended play of a finite play prefix in $\arena$ is defined analogously.

Let~$\arena$ be an arena and let $\mem_1 = (M_1, m^1_\initmark, \update_1)$ and $\mem_2 = (M_2, m^2_\initmark, \update_2)$ be memory structures for~$\arena$ and for $\arena \times \mem_1$, respectively.
We define $\mem_1 \times \mem_2 = (M_1 \times M_2, (m^1_\initmark, m^2_\initmark), \update)$, where $\update((m_1, m_2), v) = (m'_1,m'_2)$ if $\update_1(m_1, v) = m'_1$ and $\update_2(m_2, (v, m'_1)) = m'_2$.
Via a straightforward induction and in a slight abuse of notation we obtain $\update^+_{\mem_2}(\update^+_{\mem_1}(\pi)) = \update^+_{\mem_1 \times \mem_2}(\pi)$ for all finite plays~$\pi$, where we identify~$(v, m_1, m_2) = ((v, m_1), m_2) = (v, (m_1, m_2))$.

\subsection{Qualitative Games}
\label{sec:preliminaries:qualitative}

A \textit{qualitative game}~$\game = ( \arena, \wincond )$ consists of an arena $\arena$ with vertex set~$V$ and a set~$\wincond \subseteq V^\omega$ of winning plays for Player~$0$.
The set of winning plays for Player~$1$ is $V^\omega \setminus \wincond$.
As our definition of games is very general, the infinite object~$\wincond$ may not be finitely describable.
If it is, however, in a slight abuse of notation, we define~$\card{\game}$ as the sum of~$\card{\arena}$ and the size of a description of~$\wincond$.

A strategy~$\sigma$ for Player~$i$ is a \textit{winning strategy} for~$i$ in $\game = (\arena, \wincond)$ if all plays consistent with~$\sigma$ are winning for~$i$.
If Player~$i$ has a winning strategy, then we say she wins $\game$.
\textit{Solving} a game amounts to determining its winner, if one exists.
A game is determined if one player has a winning strategy.

\subsection{Quantitative Games}
\label{sec:preliminaries:quantitative}

Quantitative games extend the classical model of qualitative games.
In a quantitative game, plays are not partitioned into winning and losing plays, but rather they are assigned some measure of quality.
We keep this definition very general in order to encompass many of the already existing models.
In Section~\ref{sec:vertex-ranked-games}, we define concrete examples of such games and show how to solve them optimally.

A quantitative game~$\game = (\arena, \cost)$ consists of an arena~$\arena$ with vertex set~$V$ and a cost-function $\cost\colon V^\omega \rightarrow \nats_\infty$ for plays.
Similarly to the winning condition in the qualitative case,~$\cost$ is an infinite object.
If it is finitely describable, we, again slightly abusively, define the size~$\card{\game}$ of~$\game$ as the sum of~$\card{\arena}$ and the size of a description of~$\cost$.
A play~$\rho$ in~$\arena$ is winning for Player~$0$ in~$\game$ if~$\cost(\rho) < \infty$.%
\footnote{In order to simplify the presentation, we only consider the case in which Player~$0$ aims to minimize the cost of a play. All concepts in this work can, however, be easily adapted to the dual case in which Player~$0$ aims to maximize the cost of a play.}
Winning strategies, the winner of a game, and solving a game are defined as in the qualitative case.

We extend the cost-function over plays to strategies by defining~$\cost(\sigma) = \sup_\rho\cost(\rho)$ and~$\cost(\tau) = \inf_\rho\cost(\rho)$, where~$\rho$ ranges over the plays consistent with the strategy~$\sigma$ for Player~$0$ and over the plays consistent with the strategy~$\tau$ for Player~$1$, respectively.
Moreover, we say that a strategy~$\sigma$ for Player~$0$ (Player~$1$) is optimal if its cost is minimal (maximal) among all strategies for her (him).

For every strategy~$\sigma$ for Player~$0$, $\cost(\sigma) < \infty$ implies that~$\sigma$ is winning for Player~$0$.
However, the converse does not hold true: Each play consistent with some strategy~$\sigma$ may have finite cost, while for every~$n \in \nats$ there exists a play~$\rho$ consistent with~$\sigma$ with~$\cost(\rho) \geq n$.
Dually, a strategy~$\tau$ for Player~$1$ has~$\cost(\tau) = \infty$, if and only if~$\tau$ is winning for him.

Player~$0$ wins~$\game$ with respect to~$b$ if she has a strategy~$\sigma$ with~$\cost(\sigma) \leq b$.
Dually, if Player~$1$ has a strategy~$\tau$ with~$\cost(\tau) > b$, then he wins~$\game$ with respect to~$b$.
Solving a quantitative game~$\game$ with respect to~$b$ amounts to deciding whether or not Player~$0$ wins~$\game$ with respect to~$b$.

If Player~$0$ has a strategy~$\sigma$ with $\cost(\sigma) \leq b$, then for all strategies~$\tau$ for Player~$1$ we have~$\cost(\tau) \leq b$.
Dually, if Player~$1$ has a strategy~$\tau$ with $\cost(\tau) > b$, then for all strategies~$\sigma$ for Player~$0$ we have~$\cost(\sigma) > b$.
We say that a quantitative game is determined if for each~$b \in \nats$, either Player~$0$ has a strategy with cost at
most~$b$, or Player~$1$ has a strategy with cost strictly greater than~$b$.

We say that~$b \in \nats$ is a cap of a quantitative game~$\game$ if Player~$0$ winning~$\game$ implies that she has a strategy with cost at most~$b$.
A cap~$b$ for a game~$\game$ is tight if it is minimal.

\section{Quantitative Reductions}
\label{sec:quantitative-reductions}

\newcommand{\capfunc}{\text{cap}}

Before defining quantitative reductions, we first recall the definition of qualitative ones.
Let~$\game = (\arena, \wincond)$ and~$\game' = (\arena', \wincond')$ be qualitative games.
We say that~$\game$ is reducible to~$\game'$ via the memory structure~$\mem$ for~$\arena$ if~$\arena' = \arena \times \mem$ and if~$\rho \in \wincond$ if and only if~$\ext(\rho) \in \wincond'$.
Then, Player~$0$ wins~$\game$ if and only if she wins~$\game'$.
Moreover, if~$\sigma'$ is a winning strategy for Player~$0$ in~$\game'$ that is implemented by~$\mem'$, then a winning strategy for her in~$\game$ is implemented by~$\mem \times \mem'$.

We now define quantitative reductions as an analogous technique for the study of quantitative games and show that they exhibit the same properties as qualitative reductions.
Intuitively, given two quantitative games~$\game$ and~$\game'$, we aim to say that~$\game$ is reducible to~$\game'$ if plays in one game can be translated into plays in the other, retaining their cost along the transformation.
In fact, two such associated plays do not need to carry identical cost, but it suffices that the order on plays induced by their cost is retained.

To capture this intuition, we introduce~$b$-correction functions.
Let~$b \in \nats_\infty$.
A function~$f\colon \nats_\infty \rightarrow \nats_\infty$ is a $b$-correction function if
\begin{itemize}
	\item for all~$b'_1 < b'_2 < b$ we have~$f(b'_1) < f(b'_2)$,
	\item for all~$b' < b$ we have~$f(b') < f(b)$, and
	\item for all~$b' \geq b$ we have~$f(b') \geq f(b)$.
\end{itemize}
For $b = \infty$ these requirements degenerate to demanding that~$f$ is strictly monotonic, which in turn implies~$f(\infty) = \infty$ and~$f(b) \neq \infty$ for all~$b \neq \infty$.
Dually, if $b = 0$, we only require that~$f(0)$ bounds the values of~$f(b)$ from below.
As an example, the function~$\capfunc_b$, which is defined as~$\capfunc_b(b') = \min\set{b, b'}$ for~$b' \neq \infty$ and~$\capfunc_b(\infty) = \infty$ is a~$b$-correction function for all parameters~$b \in \nats_\infty$.

Let~$\game = (\arena,\cost)$ and~$\game' = (\arena',\cost')$ be quantitative games, let~$\mem$ be some memory structure for~$\arena$, let~$b \in \nats_\infty$, and let~$f\colon \nats_\infty \rightarrow \nats_\infty$ be some function.
We say that~$\game$ is~$b$-reducible to~$\game'$ via~$\mem$ and~$f$ if
\begin{itemize}
	\item $\arena' = \arena \times \mem$,
	\item $f$ is a $b$-correction function,
	\item $\cost'(\ext(\rho)) = f(\cost(\rho))$ for all plays~$\rho$ of~$\arena$ with~$\cost(\rho) < b$, and
	\item $\cost'(\ext(\rho)) \geq f(b)$ for all plays~$\rho$ of~$\arena$ with~$\cost(\rho) \geq b$.
\end{itemize}
We write~$\game \reducesto^b_{\mem,f} \game'$ in this case.
Moreover, we use~$\capfunc_b$ as a ``default'' function: If~$f = \capfunc_b$, we omit stating~$f$ explicitly and write~$\game \reducesto^b_\mem \game'$.
The penultimate condition implies that for each play~$\ext(\rho)$ in~$\arena'$ with~$\cost'(\ext(\rho)) \leq f(b)$ there exists some~$b'$ such that~$\cost'(\ext(\rho)) = f(b')$.

Quantitative reductions are downwards-closed with respect to the parameter~$b$:
If~$\game \reducesto^b_{\mem, f} \game'$ for some~$b \in \nats_\infty$ and~$b' \leq b$, then~$\game \reducesto^{b'}_{\mem, f} \game'$.
Moreover, similarly to the case of qualitative reductions, quantitative reductions are transitive.

\begin{thm}
\label{thm:quantitative-reductions:transitivity}
Let~$\game_1,\game_2,\game_3$ be quantitative games such that~$\game_1 \reducesto^{b_1}_{\mem_1,f_1} \game_2 \reducesto^{b_2}_{\mem_2,f_2} \game_3$ for some~$b_1, b_2 \in \nats_\infty$, some memory structures $\mem_1, \mem_2$, and some~$b_1$- and~$b_2$-correction functions $f_1$ and $f_2$.

Then, we have~$\game_1 \reducesto^{b}_{\mem, f} \game_3$, where~$\mem = \mem_1 \times \mem_2$, $f = f_2 \circ f_1$, and~$b = b_1$ if~$b_2 \geq f_1(b_1)$ and~$b = \max\set{b' \mid f_1(b') \leq b_2}$ otherwise.
\end{thm}

Quantitative reductions indeed retain the costs of strategies.
To this end, we first demonstrate that correction functions tie the cost of plays in~$\game'$ to that of plays in~$\game$.

\begin{lem}
\label{lem:cost-mapping-properties}
Let~$\game$ and~$\game'$ be quantitative games such that~$\game \reducesto^{b}_{\mem, f} \game'$, for some~$b \in \nats_\infty$, some memory structure $\mem$, and some~$b$-correction function~$f$.
All of the following hold true for all~$b' \in \nats$ and all plays~$\rho$ in~$\game$:
\begin{enumerate}
	\item\label{lem:cost-mapping-properties:1} If $b' < b$ and~$\cost'(\ext(\rho)) < f(b')$, then $\cost(\rho) < b'$.
	\item\label{lem:cost-mapping-properties:2} If $b' < b$ and~$\cost'(\ext(\rho)) = f(b')$, then $\cost(\rho) = b'$.
	\item\label{lem:cost-mapping-properties:3} If $\cost'(\ext(\rho)) \geq f(b)$, then $\cost(\rho) \geq b$.
\end{enumerate}
\end{lem}

These properties of correction functions when used in quantitative reductions enable us to state and prove the main result of this section, which establishes quantitative reductions as the quantitative counterpart to qualitative reductions:
If~$\game \reducesto^{b+1}_{\mem, f} \game'$, then all plays of cost at most~$b$ in~$\game$ are \myquot{tracked} precisely in~$\game'$.
Hence, as long as the cost of a strategy in~$\game$ is at most~$b$, it is possible to construct a strategy in~$\game'$ with cost at most~$f(b)$.
This holds true for both players.

If a strategy has cost greater than~$b$, however, we do not have a direct correspondence between costs of plays in~$\game$ and~$\game'$ anymore.
If, however,~$b$ additionally is a cap of~$\game$, and if~$\game$ is determined, then we can still show that Player~$1$ has a strategy of infinite cost in~$\game$ if he has a strategy of cost greater than~$f(b)$ in~$\game'$.

\begin{thm}
\label{thm:capped-reduction} 
Let~$\game$ and~$\game'$ be determined quantitative games such that~$\game \reducesto^{b+1}_{\mem, f} \game'$ for some~$b, \mem$, and~$f$, where~$b \in \nats$ is a cap of~$\game$.
\begin{enumerate}
	\item\label{thm:capped-reduction:exact} Let $b' < b+1$. Player~$i$ has a strategy~$\sigma'$ in~$\game'$ with $\cost'(\sigma') = f(b')$ if and only if they have a strategy~$\sigma$ in~$\game$ with $\cost(\sigma) = b'$.
	\item\label{thm:capped-reduction:cap} If Player~$1$ has a strategy~$\tau'$ in~$\game'$ with~$\cost'(\tau') \geq f(b+1)$, then he has a strategy~$\tau$ in~$\game$ with~$\cost(\tau) = \infty$.
\end{enumerate}
\end{thm}

We proved Theorem~\ref{thm:capped-reduction} by constructing optimal strategies for Player~$0$ in~$\game$ from optimal strategies for her in~$\game'$.
These strategies use the set of all play prefixes of~$\game'$ as memory states and may thus be of infinite size.
If Player~$0$ can achieve a certain cost in~$\game'$ using a finite-state strategy, however, then she can achieve the corresponding cost in~$\game$ with a finite-state strategy as well.

\begin{thm}
	\label{thm:reductions:strategy-lift}
	Let~$\game$ and~$\game'$ be quantitative games such that~$\game \reducesto^b_{\mem_1,f} \game'$ for some~$b$,~$\mem_1$, and~$f$ and let~$b' < b$.
	If Player~$i$ has a finite-state strategy~$\sigma'$ with~$\cost'(\sigma') = f(b')$ in~$\game'$ that is implemented by~$\mem_2$, then she has a finite-state strategy~$\sigma$ with~$\cost(\sigma) = b'$ in~$\game$ that is implemented by~$\mem_1 \times \mem_2$.
\end{thm}

Using quantitative reductions we are able to structure the space of quantitative games similarly to that of qualitative games.
Quantitative games are, however, usually solved by reducing them to qualitative games for a fixed bound~$b$.
Thus, there exists no ``foundation'' of the space of quantitative winning conditions, i.e., there is no canonical simple class of quantitative games that provides a natural target for quantitative reductions.
We provide such a foundation in the form of vertex-ranked games in the following section.

\section{Vertex-Ranked Games}
\label{sec:vertex-ranked-games}

\newcommand{\suprankgame}{{\textsc{Rank}^{\sup}}}
\newcommand{\limsuprankgame}{{\textsc{Rank}^{\lim}}}
\newcommand{\rankgame}{{\textsc{Rank}^{X}}}

We introduce a very simple form of quantitative games, which we call vertex-ranked games.
In such games, the cost of a play is determined solely by a qualitative winning condition and a ranking of the vertices of the arena by natural numbers.
We show that both winning conditions only incur a polynomial overhead in the complexity of solving them with respect to the underlying winning conditions.
Furthermore, we show that the memory structures implementing winning strategies for either player only incur a polynomial overhead in comparison to the memory structures implementing winning strategies for the underlying conditions.
Moreover, we discuss the optimization problem for such games, i.e., the problem of determining the minimal~$b$ such that Player~$0$ has a strategy of cost at most~$b$ in such a game.

Let~$\arena$ be an arena with vertex set~$V$, let~$\wincond \subseteq V^\omega$ be a qualitative winning condition, and let~$\rank\colon V \rightarrow \nats$ be a ranking function on vertices.
We define the quantitative vertex-ranked~$\sup$-condition
\[
	\suprankgame(\wincond, \rank)\colon v_0v_1v_2 \cdots \mapsto \begin{cases}
	 	\sup_{j \rightarrow \infty} \rank(v_j) & \text{if $v_0v_1v_2\cdots \in \wincond$} \enspace , \\
	 	\infty & \text{otherwise} \enspace , \end{cases}
\]
as well as its prefix-independent version, the vertex-ranked~$\limsup$-condition
\[ 
	\limsuprankgame(\wincond, \rank)\colon v_0v_1v_2\cdots \mapsto \begin{cases}
	 	\limsup_{j \rightarrow \infty} \rank(v_j) & \text{if $v_0v_1v_2\cdots \in \wincond$} \enspace , \\
	 	\infty & \text{otherwise} \enspace . \end{cases}
\]

A vertex-ranked $\sup$- or~$\limsup$-game~$\game = (\arena, \rankgame(\wincond, \rank))$ with~$X \in \set{\sup,\lim}$ consists of an arena~$\arena$ with vertex set~$V$, a qualitative winning condition~$\wincond$, and a vertex-ranking function~$\rank\colon V \rightarrow \nats$.

If~$\game_X = (\arena, \rankgame(\wincond, \rank))$ is a vertex-ranked~$\sup$- or~$\limsup$-game, we call the game~$(\arena, \wincond)$ the qualitative game corresponding to~$\game_X$.
Moreover, if~$\gamesup$ is a vertex-ranked~$\sup$-game, we denote the vertex-ranked~$\limsup$-game with the same arena, winning condition, and rank function by~$\gamelim$ and vice versa.
In either case, we denote the corresponding qualitative game by~$\game$.

The remainder of this section is dedicated to providing tight bounds on the complexity of solving vertex-ranked games with respect to some given bound.
In particular, we show that vertex-ranked~$\sup$-games can be solved with only an additive linear blowup compared to the complexity of solving the corresponding qualitative games.
Vertex-ranked~$\limsup$-games, on the other hand, can be solved while incurring only a polynomial blowup compared to solving the corresponding qualitative games.

\subsection{Solving Vertex-Ranked~\boldmath{$\sup$}-Games}
\label{sec:vertex-ranked-games:sup}

Let us start by noting that solving vertex-ranked $\sup$-games is at least as hard as solving the underlying qualitative games.
This is due to the fact that Player~$0$ has a winning strategy in~$(\arena, \wincond)$ if and only if she has a strategy with cost at most zero in~$(\arena, \suprankgame(\wincond, \rank))$, where~$\rank$ is the constant function assigning zero to every vertex.

\newcommand{\gamefam}{\mathfrak{G}}
\newcommand{\ver}{\text{Ver}}
\newcommand{\cpre}{\textsc{CPre}}
\newcommand{\supgames}[1]{{#1}^\textsc{rnk}_{\sup}}
\newcommand{\limsupgames}[1]{{#1}^\textsc{rnk}_{\lim}}

We now turn our attention to finding an upper bound for the complexity of solving vertex-ranked~$\sup$-games with respect to some bound.
To achieve a general treatment of such games, we first introduce some notation.
Let~$\gamefam$ be a class of qualitative games.
We define the extension of~$\gamefam$ to vertex-ranked~$\sup$-games as
\[
	\supgames{\gamefam} = \set{ (\arena, \suprankgame(\wincond, \rank)) \mid (\arena, \wincond) \in \gamefam, \rank \text{ is vertex-ranking function for } \arena } \enspace .
\]

We first show that we can use a decision procedure solving games from~$\gamefam$ to solve games from~$\supgames{\gamefam}$ with respect to a given~$b$.
To this end, we remove all vertices from which Player~$1$ can enforce a visit to a vertex of rank greater than~$b$ and proclaim that Player~$0$ wins the quantitative game with respect to~$b$ if and only if she wins the qualitative game corresponding to the resulting quantitative game.
To ensure that we are able to solve the resulting qualitative game, we assume some closure properties of~$\gamefam$.

Let~$\arena = (V, V_0, V_1, E, v_\initmark)$ and~$\arena' = (V', V'_0, V'_1, E', v'_\initmark)$ be arenas.
We say that~$\arena'$ is a \emph{sub-arena} of~$\arena$ if~$V' \subseteq V$,~$V'_0 \subseteq V_0$,~$V'_1 \subseteq V_1$, and~$E' \subseteq E$ and write~$\arena' \subarenaof \arena$ in this case.
We call a class of qualitative (or quantitative) games~$\gamefam$ \emph{proper} if for each~$(\arena, \wincond)$ (or $(\arena, \cost)$) in $\gamefam$ and each sub-arena~$\arena' \subarenaof \arena$ the game~$(\arena', \wincond')$ (or $(\arena', \cost')$), where~$\wincond'$ (or~$\cost'$) is the restriction of~$\wincond$ (or~$\cost$) to vertices from~$\arena'$, is a member of~$\gamefam$ as well, if all games in $\gamefam$ are determined and if all~$\game \in \gamefam$ are finitely representable.

Moreover, in order to formalize the idea of removing vertices from which one player can enforce a visit to some set of vertices, we recall the attractor construction.
Let~$\arena = (V, V_0, V_1, E, v_\initmark)$ be an arena with~$n$ vertices and let~$X \subseteq V$.
We define~$\att{i}{}{X} = \att{i}{n}{X}$ inductively with~$\att{i}{0}{X} = X$ and
\begin{multline*}
	\att{i}{j}{X} = \set{ v \in V_i \mid \exists v' \in \att{i}{j-1}{X}.\, (v,v') \in E} \, \cup \\
		\set{ v \in V_{1-i} \mid \forall (v, v') \in E.\, v' \in \att{i}{j-1}{X}} \cup \att{i}{j-1}{X} \enspace.
\end{multline*}
Intuitively, the $i$-attractor $\att{i}{}{X}$ is the set of all vertices from which Player~$i$ can enforce a visit to~$X$.
The set~$\att{i}{}{X}$ can be computed in linear time in~$\card{E}$ and Player~$i$ has a positional strategy~$\sigma$ such that each play starting in some vertex in $\att{i}{}{X}$ and consistent with~$\sigma$ eventually encounters some vertex from~$X$~\cite{NerodeRemmelYakhnis96}.
We call~$\sigma$ the attractor strategy towards~$X$.

Player~$0$ wins~$\gamesup \in \supgames{\gamefam}$ with respect to some bound~$b$ if and only if the initial vertex of~$\gamesup$ is not in the $1$-attractor~$A$ of the vertices of rank greater than~$b$, and if she is able to win the game~$\game'$ obtained from~$\game$ by removing the $1$-attractor of~$A$.
Since~$A$ is computable in linear time, we obtain a generic decision procedure solving vertex-ranked~$\sup$-games in linear time and space.

\begin{thm}
\label{thm:vertex-ranked-games:direct:complexity}
Let~$\gamefam$ be a proper class of qualitative games~$\game$ that can be solved in time~$t(\card{\game})$ and space~$s(\card{\game})$, where~$t$ and~$s$ are monotonic functions.

Then, the following problem can be solved in time~$\bigo(n) + t(\card{\game})$ and space $\bigo(n) + s(\card{\game})$:
\myquot{Given some game~$\gamesup \in \supgames{\gamefam}$ with~$n$ vertices and some bound~$b \in \nats$, does Player~$0$ win~$\gamesup$ with respect to~$b$?}
\end{thm}

This theorem provides an upper bound on the complexity of solving vertex-ranked~$\sup$-games.
Intuitively, we prove Theorem~\ref{thm:vertex-ranked-games:direct:complexity} by showing that, for any vertex-ranked~$\sup$-game~$\gamesup$, a winning strategy for Player~$0$ in~$\game$ that never moves to the Player~$1$-attractor towards vertices of rank greater than~$b$ has cost at most~$b$.
Thus, an upper bound on the size of winning strategies for Player~$0$ for games from~$\gamefam$ provides an upper bound for strategies of finite cost in~$\supgames{\gamefam}$ as well.
Moreover, if the decision procedure deciding~$\gamefam$ constructs winning strategies for one or both players, we can adapt the decision procedure deciding~$\supgames{\gamefam}$ to construct strategies of cost at most (greater than)~$b$ for Player~$0$ (Player~$1$) as well.

\begin{cor}
	Let~$\gamefam$ be a proper class of qualitative games and let~$\game \in \gamefam$.
	If $\sigma$ is a finite-state winning strategy for Player~$i$ in~$\game$, then Player~$i$ has a finite-state winning strategy~$\sigma_{\sup}$ in~$\gamesup$ with~$\card{\sigma_{\sup}} \in \bigo(\card{\sigma})$.
	Furthermore, if~$\sigma$ is effectively constructible, then~$\sigma_{\sup}$ is effectively constructible.
\end{cor}

Finally, this procedure enables us to solve the optimization problem for vertex-ranked $\sup$-games from~$\supgames{\gamefam}$:
Recall that if Player~$0$ wins~$\gamesup$ with respect to some~$b$, she wins it with respect to all~$b' \geq b$ as well.
Hence, using a binary search,~$\log(n)$ invocations of the decision procedure obtained in the proof of Theorem~\ref{thm:vertex-ranked-games:direct:complexity} suffice to determine the minimal~$b$ such that Player~$0$ wins~$\gamesup$  with respect to~$b$.
Hence, it is possible to determine the minimal such~$b$ in time $\bigo(\log(n)(n + t(\card{\game})))$ and space~$\bigo(n) + s(\card{\game})$.

\subsection{Solving Vertex-Ranked~\boldmath$\limsup$-Games}
\label{sec:vertex-ranked-games:limsup}

We now turn our attention to solving vertex-ranked~$\limsup$-games.
Solving these games is again at least as hard as solving their corresponding qualitative games, due to the same reasoning as given above for vertex-ranked~$\sup$-games.
Thus, we again only provide upper bounds on the complexity of solving such games.
To this end, given some class~$\gamefam$ of games, we define the corresponding class of vertex-ranked~$\limsup$-games
\[
	\limsupgames{\gamefam} = \set{ (\arena, \limsuprankgame(\wincond, \rank)) \mid (\arena, \wincond) \in \gamefam, \rank \text{ is vertex-ranking function for } \arena } \enspace .
\]

\newcommand{\cobuchi}{\textsc{CoBüchi}}

We identify two criteria on classes of qualitative games~$\gamefam$, each of which is sufficient for quantitative games in~$\limsupgames{\gamefam}$ to be solvable with respect to some given~$b$.
More precisely, we provide decision procedures for~$\limsupgames{\gamefam}$ for the case that
\begin{enumerate}
	\item games from~$\gamefam$ can be solved in conjunction with coBüchi-conditions, and for the case that
	\item games from~$\gamefam$ are prefix-independent.
\end{enumerate}

For the first case, fix some class of games~$\gamefam$ and let~$\gamelim = (\arena, \limsuprankgame(\wincond, \rank)) \in \limsupgames{\gamefam}$ with vertex set~$V$.
Moreover, recall that a play in~$\wincond$ has cost at most~$b$ in~$\gamelim$ if it visits vertices of rank greater than~$b$ only finitely often.
In general, such behavior is formalized by the qualitative co-Büchi condition~$\cobuchi(F) = \set{\rho \in V^\omega \mid \inf(\rho) \cap F = \emptyset}$, where~$\inf(\rho)$ denotes the set of vertices occurring infinitely often in~$\rho$.
Clearly, Player~$0$ has a strategy of cost at most~$b$ in~$\gamelim$ if and only if she wins~$(\arena, \wincond \cap \cobuchi(\set{v \in V \mid \rank(v) > b}))$.
This observation gives rise to the following remark.

\begin{rem}
Let~$\gamefam$ be a class of qualitative games such that the games in~$\set{(\arena, \wincond \cap \cobuchi(F)) \mid (\arena, \wincond) \in \gamefam, F \subseteq V, V \text{ is vertex set of } \arena}$ can be solved in time~$t(\card{\game}, \card{F})$ and space~$s(\card{\game}, \card{F})$, where~$t$ and~$s$ are monotonic functions.

Then, the following problem can be solved in time~$t(\card{\gamelim}, n)$ and space $s(\card{\gamelim}, n)$:
\myquot{Given some game~$\gamelim \in \limsupgames{\gamefam}$ with~$n$ vertices and some bound~$b \in \nats$, does Player~$0$ win~$\gamelim$ with respect to~$b$?}
\end{rem}

In this case, we solve vertex-ranked~$\limsup$-games via a decision procedure for solving qualitative games as-is.
Such a procedure trivially exists if the winning conditions of games from~$\gamefam$ are closed under intersection with co-Büchi conditions.
Thus, we obtain solvability of a wide range of classes of vertex-ranked~$\limsup$-games, e.g., co-Büchi-, parity-, Muller-, Streett- and Rabin games.

We now turn our attention to the second case described above:
We consider classes~$\limsupgames{\gamefam}$ where a play is only determined to be winning or losing in a game from~$\gamefam$ due to some infinite suffix.
Formally, we say that a qualitative winning condition~$\wincond \subseteq V^\omega$ is \emph{prefix-independent} if for all infinite plays~$\rho \in V^\omega$ and all play prefixes~$\pi \in V^*$, we have $\rho \in \wincond$ if and only if~$\pi\rho \in \wincond$.
A qualitative game is prefix-independent if its winning condition is prefix-independent.
A class of games is prefix-independent if every game in the class is prefix-independent.

Let~$\gamefam$ be a prefix-independent class of games and let~$\gamelim \in \limsupgames{\gamefam}$.
Moreover, let~$b \in \nats$.
In order to solve~$\gamelim$ with respect to~$b$, we adapt the classic algorithm for solving prefix-independent qualitative games (cf., e.g., the work by Chatterjee, Henzinger, and Piterman~\cite{ChatterjeeHenzingerPiterman06}).
Thereby, we repeatedly compute the set of vertices from which Player~$0$ has a strategy of cost at most~$b$ in the corresponding vertex-ranked~$\sup$-game ~$\gamesup$ and remove their attractor from the game similarly to the construction of a decision procedure for vertex-ranked $\sup$-games in Theorem~\ref{thm:vertex-ranked-games:direct:complexity}.
We claim that Player~$0$ has a strategy with cost at most~$b$ in~$\gamelim$ if and only if~$v_\initmark$ was not removed during the above construction.

\begin{thm}
\label{thm:vertex-ranked-games:indirect:complexity}
Let~$\gamefam$ be a proper prefix-independent class of qualitative games~$\game$ that can be solved in time~$t(\card{\game})$ and space~$s(\card{\game})$, where~$t$ and~$s$ are monotonic functions.

Then, the following problem can be solved in time~$\bigo(n^3 + n^2\cdot t(\card{\gamelim}))$ and space $\bigo(n + s(\card{\gamelim}))$:
\myquot{Given some game~$\gamelim \in \limsupgames{\gamefam}$ with~$n$ vertices and some bound~$b \in \nats$, does Player~$0$ win~$\gamelim$ with respect to~$b$?}
\end{thm}

Intuitively, we prove Theorem~\ref{thm:vertex-ranked-games:indirect:complexity} by constructing a strategy~$\sigma$ for Player~$0$ by \myquot{stitching together} the attractor-strategies towards her winning regions in the decreasing vertex-ranked~$\sup$-games and the winning strategies for her in the respective vertex-ranked~$\sup$-games.
As each play consistent with that strategy descends down the hierarchy of~$\sup$-games thus constructed, we can reuse the memory states of the winning strategies in these games when implementing~$\sigma$.
Thus, a monotonic upper bound on the size of strategies with cost at most~$b$ in~$\gamesup$ is an upper bound on the size of such strategies in~$\gamelim$ as well.

\begin{cor}
Let~$\gamefam$ be a proper prefix-independent class of qualitative games such that, if Player~$0$ wins~$\game$, then she has a finite-state winning strategy of size at most~$m(\card{\game})$, where~$m$ is a monotonic function.

If Player~$0$ wins~$\gamelim \in \limsupgames{\gamefam}$ with finite-state strategies, then she has a finite-state winning strategy~$\sigma_{\lim}$ of size at most~$m(\card{\gamelim})$ in~$\gamelim$.	
Furthermore, if winning strategies for Player~$i$ in the games in~$\gamefam$ are effectively constructible, then~$\sigma_{\lim}$ is effectively constructible.
\end{cor}

Moreover, in order to find the optimal~$b$ such that Player~$0$ wins~$\gamelim$ with respect to~$b$, we can again employ a binary search.
Thus, we can determine the optimal such~$b$ in time~$\bigo(\log(n)(n^3 + n^2\cdot t(\card{\game})))$ and space $\bigo(n + s(\card{\game}))$.

Having thus defined both quantitative reductions as well as a canonical target for such reductions, we now give an example of how to solve quantitative games using this tools.

\section{Applications}
\label{sec:application}

In this section, we give examples of how to use quantitative reductions and vertex-ranked games to solve quantitative games.
First, in Section~\ref{sec:applications:request-response}, we formally introduce a quantitative variant of request-response games, which we call request-response games with costs, and show how to solve such games using quantitative reductions and vertex-ranked sup-request-response games.
Subsequently, in Section~\ref{sec:application:resilience}, we show that vertex-ranked games are useful in their own right, by showing how to use them to synthesize controllers that are resilient against disturbances.

\subsection{Reducing Request-Response Games with Costs to Vertex-Ranked Request-Response Games}
\label{sec:applications:request-response}

Recall that a play satisfies the qualitative request-response condition if every request that is opened is eventually answered.
We extend this condition to a quantitative one by equipping the edges of the arena with costs and measuring the maximal cost incurred between opening and answering a request. 

Fix some arena~$\arena$ with vertex set~$V$ and set~$E$ of edges.
Formally, the qualitative request-response condition~$\reqres(\Gamma)$ consists of a family of so-called request-response pairs~$\Gamma = (Q_c, P_c)_{c \in [d]}$, where $Q_c,P_c \subseteq V$ for all~$c \in [d]$.
Player~$0$ wins a play according to this condition if each visit to some vertex from~$Q_c$ is answered by some later visit to a vertex from~$P_c$, i.e., we define
\[
	\reqres((Q_c, P_c)_{c \in [d]}) = \set{ v_0v_1v_2\cdots \in V^\omega \mid \forall c \in [d] \forall j \in \nats.\, v_j \in Q_c \text{ implies } \exists j' \geq j.\, v_{j'} \in P_c} \enspace.
\]
We say that a visit to a vertex from~$Q_c$ \emph{opens a request for condition}~$c$ and that the first visit to a vertex from~$P_c$ afterwards \emph{answers the request for that condition}.

\begin{prop}[\cite{WallmeierHuettenThomas03}]\label{thm:request-response:prior-work}
Request-response games with~$n$ vertices and~$d$ request-response pairs can be solved in time $\bigo(n^2 d^2 2^d)$.

Furthermore, let~$\game$ be a request-response game with~$d$ request response pairs.
If Player~$0$ has a winning strategy in~$\game$, then she has a finite-state winning strategy of size at most~$d2^d$.
\end{prop}

We extend this winning condition to a quantitative one using families of cost functions~$\cost = (\cost_c)_{c \in [d]}$, where~$\cost_c\colon E \rightarrow \nats$ for each~$c \in [d]$ and lift the cost functions~$\cost_c$ to play infixes~$\pi$ in~$\arena$ by adding up the costs along~$\pi$.
The cost-of-response for a request for condition~$c$ at position~$j$ is then defined as
\[
	\reqresdist_c(v_0v_1v_2\cdots, j) =
	\begin{cases}
 		\min \set{ \cost_c (v_j \cdots v_{j'}) \mid j' \ge j \text{ and } v_{j'} \in P_c } &\text{if } v_j \in Q_c \enspace , \\
 		0 &\text{otherwise} \enspace ,
 	\end{cases}
\]
with~$\min \emptyset = \infty$, which naturally extends to the (total) cost-of-response
\[
	\reqresdist(\rho, j) = \max\nolimits_{c \in [d]} \reqresdist_c(\rho, j) \enspace.
\]

Finally, we define the request-response condition with costs as 
\[
	\costreqres(\Gamma, \cost)(\rho) = \sup\nolimits_{j\rightarrow \infty} \reqresdist(\rho , j) \enspace,
\]
i.e., it measures the maximal cost incurred by any request in~$\rho$.

We call a game~$\game = (\arena, \costreqres(\Gamma, \cost))$ a request-response game with costs.
We denote the largest cost assigned to any edge by~$W$.
As we assume the functions~$\cost_c$ to be given in binary encoding, the largest cost~$W$ assigned to an edge may be exponential in the description length of~$\game$.

If all~$\cost_c$ assign zero to every edge, then the request-response condition with costs coincides with the qualitative request-response condition.
In general, however, the request-response condition with costs is a strengthening of the classical request-response condition:
If some play~$\rho$ has finite cost according to the condition with costs, then it is winning for Player~$0$ according to the qualitative condition, but not vice versa.

\begin{rem}
\label{rem:request-response:cost-to-classical}
Let~$\game = (\arena, \costreqres(\Gamma, \cost))$ be a request-response game with costs.
If a strategy~$\sigma$ for Player~$0$ in~$\game$ has finite cost, then~$\sigma$ is a winning strategy for Player~$0$ in~$(\arena, \reqres(\Gamma))$.
\end{rem}

This remark together with a detour via qualitative request-response games yield a cap for request-response games with costs.

\begin{lem}
\label{lem:request-response:cap}
Let~$\game$ be a request-response game with costs with~$n$ vertices,~$d$ request-response pairs, and largest cost of an edge~$W$.
If Player~$0$ has a strategy with finite cost in~$\game$, then she also has a strategy with cost at most $d 2^d n W$.
\end{lem}

Having obtained a cap for request-response games with costs, we can now turn to the main result of this section:
Request-response games with costs are reducible to vertex-ranked $\sup$-request-response games.
In order to show this, we use a memory structure that keeps track of the costs incurred by the requests open at each point in the play~\cite{WeinertZimmermann16}.

\begin{lem}
\label{lem:request-response:reduction}
	Let~$\game$ be a request-response game with costs with~$n$ vertices,~$d$ request-response pairs, and highest cost of an edge~$W$.
	Then~$\game \reducesto^{b+1}_\mem \game'$ for $b = d2^d nW$, some memory structure~$\mem$ of size~$\bigo(2nb^d)$, and a vertex-ranked~$\sup$-request-response game~$\game'$ with~$d$ request-response pairs.
\end{lem}

Thus, in order to solve a request-response game with costs with respect to some~$b$, it suffices to solve a vertex-ranked $\sup$-request-response game with respect to~$b$.
This, in turn, can be done by reducing the problem to that of solving a request-response game as shown in Theorem~\ref{thm:vertex-ranked-games:direct:complexity}.
Using this reduction together with the framework of quality-preserving reductions, we are able to provide an upper bound on the complexity of solving request-response games with respect to some bound~$b$.

\begin{thm}
	\label{thm:request-response:exptime-membership}
	The following decision problem is in \exptime:
	\myquot{Given some request-response game with costs~$\game$ and some bound~$b \in \nats$, does Player~$0$ have a strategy~$\sigma$ with~$\cost(\sigma) \leq b$ in~$\game$?}
\end{thm}

Moreover, solving request-response games is known to be \exptime-hard~\cite{ChatterjeeHenzingerHorn11}.
Thus, solving quantitative request-response games via quantitative reductions is asymptotically optimal.

Also, recall that Player~$0$ has a strategy with cost~$b'$ in some request-response game with costs if and only if she has a strategy with cost~$b'$ in the vertex-ranked $\sup$-request-response game~$\game'$ constructed in the proof of Lemma~\ref{lem:request-response:reduction}, which has as many request-response pairs~$d$ as~$\game$.
Due to Proposition~\ref{thm:request-response:prior-work}, if she has a strategy of cost at most~$b'$ in~$\game'$, she has one of the same cost and of size at most~$d2^d$ in~$\game'$, as argued in Section~\ref{sec:vertex-ranked-games:sup}.
Hence, due to Theorem~\ref{thm:reductions:strategy-lift}, we obtain an exponential upper bound on the size of optimal strategies for Player~$0$.

\begin{cor}
Let~$\game$ be a request-response game with costs with~$n$ vertices,~$d$ request-response pairs, and highest cost of an edge~$W$.
If Player~$0$ has a strategy with finite cost, then she also has a strategy with the same cost of size at most~$\bigo(n b^d d 2^d)$, where~$b = d 2^d n W$.
\end{cor}

Finally, the optimization problem of finding the minimal~$b'$ such that Player~$0$ wins a request-response game~$\game$ with respect to~$b'$ can be solved in exponential time as well.
Recall that if Player~$0$ wins~$\game$ with respect to some~$b'$, then she also wins it with respect to all~$b'' \geq b'$.
Since we can assume~$b' \leq b = d 2^d n W$, we can perform a binary search for~$b'$ on the interval $\set{0,\dots,b}$.
Hence, the optimal~$b'$ can be found in time~$\bigo(\log(b) n^3 b^{3d} d 2^d)$.

\subsection{Fault Resilient Strategies for Safety Games}
\label{sec:application:resilience}

We now demonstrate the flexibility and versatility of vertex-ranked games in their own right.
To this end, we consider the problem of synthesizing a controller for a reactive system that is embedded into some environment.
This setting is typically modeled as an infinite game in which Player~$0$ and Player~$1$ take the roles of the controller and of the environment, respectively.
Here, we consider safety games, i.e., we assume that the specification for the controller is given as a game in which it is the aim of Player~$0$ to keep the play inside a safe subset of the vertices.

Dallal et al.~\cite{DallalNeiderTabuada16} argue that this setting is not sufficiently expressive to correctly model a real-world scenario, since it assumes that Player~$0$ can accurately predict the effect of her actions on the state of the system.
In a realistic setting, in contrast, faults may occur, i.e., an action chosen by a controller may be executed incorrectly, or it may not be executed at all.

In order to model such faults, Dallal et al\ introduce arenas with faults~$\arena_F = (V, V_0, V_1, E, F, v_\initmark)$, which consist of an arena~$(V, V_0, V_1, E, v_\initmark)$ and a set of faults~$F \subseteq V_0 \times V$.
In such an arena, whenever it is Player~$0$'s turn, say at vertex~$v$, a fault $(v, v') \in F$ may occur, resulting in the play continuing in vertex~$v'$ instead of that chosen by Player~$0$.
Moreover, Dallal et al.\ consider safety conditions, i.e., qualitative winning conditions of the form~$\safety(S)= \set{ v_0v_1v_2\cdots \mid \forall i \in \nats.\, v_i \in S}$ for some~$S \subseteq V$.
Hence, it is the aim of Player~$0$ to keep the play inside the ``safe'' set of vertices~$S$.
If the play leaves the set~$S$, it is declared winning for Player~$1$.
The task at hand is to compute a fault-resilient strategy for Player~$0$ that forces the play to remain inside~$S$ and that can ``tolerate'' as many faults as possible.

Safety games without faults are solved by a simple attractor construction:
As soon as the play enters~$W_1 = \att{1}{}{V \setminus S}$, Player~$1$ can play consistently with his attractor strategy towards~$V \setminus S$ in order to win the play.
Thus, it is Player~$0$'s aim to keep the play inside $W_0 = V \setminus W_1$.

Dallal et al.\ solve the problem of computing fault-resilient strategies for safety games by adapting the classical algorithm for solving safety games to this setting.
In doing so, they obtain a value~$\val(v)$ for each vertex~$v$ that denotes the minimal number of faults that need to occur in order for the play to reach~$W_1$, if Player~$0$ plays well.
Furthermore, they show that~$\val$ can be computed in polynomial time in~$\card{V}$.
Finally, due to the existence of positional winning strategies for both players in safety games, they obtain~$\val(v) \in [n] \cup \set{\infty}$ for all~$v \in V$.
Then, a fault-resilient strategy for Player~$0$ is one that maximizes the minimal value~$\val(v)$ witnessed during any play.
Dallal et al.\ construct such a strategy on the fly during the computation of~$\val(v)$~\cite{DallalNeiderTabuada16}.

This task can, however, easily be reframed as a vertex ranked game in the arena $\arena = (V, V_0, V_1, E, v_\initmark)$, which we obtain from~$\arena_F$ by omitting the faults.
In that game, we assign to each vertex the rank~$\rank(v) = \card{V} - \val(v)$ if~$\val(v) \in [n]$ and~$\rank(v) = 0$ otherwise, i.e., if~$\val(v) = \infty$.
Then, Player~$0$ has a strategy with cost at most~$b$ in~$\game' = (\arena, \suprankgame(\safety(S), \rank))$ if and only if she has a winning strategy in the original safety game with faults that tolerates at least~$\card{V} - b$ faults.

This formulation as a vertex-ranked game enables further study of games in arenas with faults.
Here, we require the winning condition to be a safety condition in order to compute~$\val(v)$.
In recent work, we have shown how to compute this value for more complex qualitative winning conditions~\cite{NeiderWeinertZimmermann18}.
If~$\val(v)$ is effectively computable for a given qualitative winning condition, one can easily obtain fault-resilient strategies by formulating the task as a vertex-ranked game as demonstrated.

Finally, the formulation as a vertex-ranked game yields a method to compute eventually-fault-resilient strategies, i.e., strategies that are resilient to a large number of faults after a finite ``start-up'' phase.
In order to obtain such strategies, it suffices to view the resulting vertex-ranked games as a $\limsup$-game instead of a~$\sup$-game and to solve it optimally as described in Section~\ref{sec:vertex-ranked-games:limsup}.

\section{Conclusion}

In this work, we have lifted the concept of reductions, which has yielded a multitude of results in the area of qualitative games, to quantitative games.
We have shown that this novel concept exhibits the same useful properties for quantitative games as it does for qualitative ones and that it furthermore retains the quality of strategies.

Additionally, we have provided two very general types of quantitative games that serve as targets for quantitative reductions, namely vertex-ranked~$\sup$ games and vertex-ranked~$\limsup$-games.
For both kinds of games we have shown a polynomial overhead on the complexity of solving them with respect to some bound, on the memory necessary to achieve a given cost, and on the complexity of determining the optimal cost that either player can ensure.

Finally, we have demonstrated the versatility of these tools by using them to solve quantitative request-response games and by showing how to solve the problem of computing fault-resilient strategies in safety games via vertex-ranked games.
This latter formulation enables a general study of games with faults, even in the presence of more complex winning conditions than the safety condition considered by Dallal et al.~\cite{DallalNeiderTabuada16} and in this work.
We are currently investigating how to leverage vertex-ranked games for the synthesis of fault-resistant strategies in parity games.

Further research continues in two additional directions:
Firstly, while the framework of quantitative reductions and vertex-ranked games yields upper bounds on the complexity of solving quantitative games, it does not directly yield lower bounds on the complexity of the problems under investigation.
Consider, for example, the problem of solving parity games with costs with respect to some bound, which is~$\pspace$-complete~\cite{WeinertZimmermann16}.
It is possible to reduce this problem to that of solving a vertex-ranked parity game of exponential size and linearly many colors similarly to the reduction presented in this work, which yields an~$\exptime$-algorithm.
It remains open how to use quantitative reductions to obtain an algorithm for this problem that only requires polynomial space.

Secondly, another goal for future work is the establishment of an analogue to the Borel hierarchy for quantitative winning conditions.
In the qualitative case, this hierarchy establishes clear boundaries for reductions between infinite games, i.e., a game whose winning condition is in one level of the Borel hierarchy cannot be reduced to one with a winning condition in a lower level.
Also, each game with a winning condition in the hierarchy is known to be determined~\cite{Martin75}.
To the best of our knowledge, it is open how to define such a hierarchy for quantitative winning conditions which exhibit similar properties.

\subparagraph*{Acknowledgements}
I would like to thank Martin Zimmermann for many fruitful discussions.

\bibliographystyle{eptcs}
\bibliography{main}

\end{document}